# The relationship of electron Fermi energy with strong magnetic fields

Qiu He Peng[a](E-mail: qhpeng@nju.edu.cn) , Zhi Fu Gao[b] and  Na Wang[b]
**a .**Department of Astronomy, Nanjing University, Nanjing, 210093, China
**b.** Xinjiang Astronomical Observatory, Chinese Academy of Sciences, Urumqi, 830011, China

**Abstract:**  In order to depict the quantization of Landau levels, we introduce Dirac $\delta$ -function, and gain a concise expression for the electron Fermi energy , $E_F(e) \propto B^{1/4}$. The high soft  X-ray luminosities of magnetars  may be naturally explained by our theory .

## I. Two contrary views on the relationship between $E_F(e)$ and $\mathcal{B}$

In the light of the viewpoint on the quantization of Landau levels, the movements of free electrons are continuous along the magnetic field, whereas the energies of free electrons moving in the direction perpendicular to the magnetic field are quantized or discrete.  In an intense magnetic field, the  energy of an electron  can be expressed as follows:

$$(\frac{E_e(p_z,b,n,\sigma)}{m_e c^2})^2 = 1 + (\frac{p_z}{m_e c})^2 + 2(n+\tfrac{1}{2}+\sigma)\frac{2\mu_e B}{m_e c^2} = 1 + (\frac{p_z}{m_e c})^2 + 2(n+\tfrac{1}{2}+\sigma)b, \quad (1)$$

where, $p_z$ and $p_\perp$ are the electron momentum along the magnetic field  and the electron momentum perpendicular to the magnetic field, respectively; $n = 0,1,2,3,\cdots$ is the Landau  level quantum number; $\sigma$ is the spin quantum number of an electron, $\sigma = -1/2$ when $n = 0$ ; $\sigma = \pm 1/2$ when $n \geq 1$, $B_{cr} = m_e^2 c^3 / e\hbar = 4.414 \times 10^{13}$ is the critical magnetic field[1]. In a weak magnetic field ( $B \ll B_{cr}$ ), both $p_z$ and $p_\perp$ change continuously, and the Fermi surface  (in a momentum space) of electrons is spherically symmetric basically.  In quantum statistics, $h^3$  is a fundamental unit of phase- space volume. The phase-space element in the volume element  $d^3 x$ is $d^3 x d^3 p$ ,  so the number of microscopic state in $d^3 x d^3 p$  is $d^3 x d^3 p / h^3$, where $h$ is Planck's constant, The Fermi sphere, however,  transforms into a Landau column in an intense magnetic field.  In the x-y plane, electrons are populated on discrete Landau levels, though $p_z$ changes continuously.  For a given $p_z$ ( $p_z$ is still continuous)，there is a maximum orbital quantum number ( or Landau level number ) $n_{\max}(p_z,b,\sigma)$. When $B \gg B_{cr}$ or $b \gg 1$, the Landau column becomes a very long , narrow  cylinder along the magnetic field, and  the overwhelming majority of electrons congregate in the lowest levels with $n_L = 0$, 1.  Due to a given number density of electrons in the interior of a neutron star,  the stronger the magnetic field, the fewer the number of  states occupied by electrons in the direction perpendicular to the magnetic field, according to the Pauli exclusion principle (each microscopic state is occupied by one electron only). Therefore, the maximum of momentum along the magnetic field $p_z(\max)$ will increase with increasing magnetic field, and the electron  Fermi energy will also increase, where the relation $p_z(\max) \approx p_F \approx E_F(e)/c$ for ultra-relativistic electrons is used[1,2]. Then we naturally draw a conclusion that the stronger the magnetic field, the higher the Fermi energy of electrons. Unfortunately, we are surprised to find that currently the most popular viewpoint --" the stronger the magnetic field, the lower the electron Fermi energy" , is completely contrary to our conclusion above. Most seriously, this popular view has appeared in many papers on astrophysics, e.g. refs.[3-5]. We propose a correct expression for the electron Fermi energy in super strong magnetic fields in the paper firstly. Furthermore, our theory can provide a reasonable explanation of an important astronomical phenomena that magnetars universally possess high  soft  X-ray  luminosities.

## II. The key to the question

Inside a neutron star including the crust, electrons are relativistic and degenerate. According to statistical physics, the microscopic state number in a 6-dimension phase-space element $dxdydzdp_xdp_ydp_z$ is $dxdydzdp_xdp_ydp_z/h^3$. In the Ref.[1], the number of states occupied by completely degenerate relativistic electrons in an unit volume is calculated as follows:

$$N_{phase} = \sum_{p_z}\sum_{p_y}\sum_{p_x} = \frac{1}{h^3}\int_{-\infty}^{\infty}\int_{-\infty}^{\infty}\int_{-\infty}^{\infty}dp_xdp_ydp_z = \frac{1}{h^3}\int_0^{p_F}dp_z\int_0^{\infty}p_\perp dp_\perp\int_0^{2\pi}d\Phi = \frac{\pi p_F}{h^3}\int_0^{\infty}dp_\perp^2, \quad (2)$$

where $\Phi = \tan^{-1}p_y/p_x$. Quantization requires $p_\perp^2 \to m_e^2c^4\frac{B}{B_{cr}}2n$, hence $\int_0^{\infty}dp_\perp^2 \to \sum_{n=0}^{\infty}\omega_n$, $\omega_n$ is the degeneracy of the n-th level, which can be estimated as

$$\omega_n = \frac{1}{h^2}\int_0^{2\pi}d\Phi\int_{A<p_\perp^2<B}p_\perp dp_\perp = \frac{2\pi}{h^2}\frac{(B-A)}{2} = \frac{1}{2\pi}(\frac{\hbar}{m_ec})^{-2}\frac{B}{B_{cr}}, \quad (3)$$

$A = m_e^2c^2\frac{2nB}{B_{cr}}$, $B = m_e^2c^2\frac{2(n+1)B}{B_{cr}}$. Note, the above calculation method was cited from S*tatistical Mechanics,1965*[6]. The classical textbook, S*tatistical Mechanics,2003* [7], also gave

$$\frac{1}{h^2}\int dp_xdp_y = \frac{1}{h^2}\pi p_\perp^2\Big|_n^{n+1} = \frac{4\pi m_e\mu_e B}{h^2} = \frac{1}{2\pi}(\frac{\hbar}{m_ec})^{-2}\frac{B}{B_{cr}}, \quad \mu_e = \frac{e\hbar}{2m_ec} \quad . (4)$$

From Eq.(3) and Eq.(4), the expression for the degeneracy of the relativistic landau level is in complete accordance with the corresponding nonrelativistic case[8], that is rare. Eq.(3) or Eq.(4) is just the expression we are looking for, that leads to an incorrect viewpoint on the electron Fermi energy prevailing in the world currently. In the interior of a neutron star, in the light of the Pauli exclusion principle, the electron number density should be equal to its microscopic state density, , $n_e = N_{phase} = N_A\rho Y_e$ where $N_A$ is the Avogadro constant, $\rho$ is the matter density, and $Y_e$ is the electron fraction. From the analysis above, it is easy to gain $E_F(e) \propto B^{-1}$, that is, the electron Fermi energy decreases with increasing magnetic field. However, after careful consideration and analysis, in our idea, the above expression for $\omega_n$ is incorrect, because it's against the viewpoint on the quantization of Landau levels. The essence of the above method (or the incorrect deduction) lies in the assumption that the torus located between the n-th Landau level and the (n+1)-th Landau level in momentum space is ascribed to the (n +1)-th Landau level. Such a factitious assumption is equivalent to allow a continuous momentum (or energy) of an electron moving in the direction perpendicular to the magnetic field, which is obviously contradictory to the quantization of Landau levels in the case of superhigh magnetic fields. The concept of the Landau level quantization clearly tells us that there is no any microscopic quantum state between $p_\perp(n_n)$ and $p_\perp(n_{n+1})$. In a word, the main cause of the popular incorrect view on $E_F(e)$ is due to a factitious assumption. In order to depict the quantization of landau Levels truly and accurately, we must introduce the Dirac $\delta-$function. As is well known, the eigenvector wave function of the Schrodinger equation (or Dirac equation) can be expanded in an infinite series. In the process of deducing the expressions concerning the quantization of Landau levels, we should firstly give a $p_z$ that changes continuously along the magnetic field, then solve the relativistic Schrodinger equation (or Dirac equation), $n_{max}$, the maximum of Landau level quantum number, is obtained by truncating the infinite series when the wave function is limited[8]. Logically, give a $p_z$ first and then determine the maximum Landau level number, $n_{max}$,

$$n_{max}(\sigma = \frac{1}{2}) = Int\{\frac{1}{2b}[(\frac{E_F(e)}{m_ec^2})^2 - 1 - (\frac{p_z}{m_ec})^2] - 1\}, n_{max}(\sigma = \frac{-1}{2}) = Int\{\frac{1}{2b}[(\frac{E_F(e)}{m_ec^2})^2 - 1 - (\frac{p_z}{m_ec})^2]\}$$

As an alternative way to depict the quantization of landau Levels, starting from here, we make use of the Dirac $\delta-$ function $\delta(\frac{p_\perp}{m_e c}-[2(n+\frac{1}{2}+\sigma)b]^{1/2})$, then Eq.(2) should be rewritten as

$$N_{phase} = 2\pi \left(\frac{m_e c}{h}\right)^3 \int_0^{E_F/m_e c} d\frac{p_z}{m_e c} \sum_{n=0}^{n_{max}(p_z,b,\sigma)} g_n \int_0^{E_F/mc} \delta(\frac{p_\perp}{m_e c}-\sqrt{2(n+\frac{1}{2}+)b})\frac{p_\perp}{m_e c} d\frac{p_\perp}{m_e c}, \quad (5)$$

where $g_n$ is the spin degeneracy, $g_n=1$ for n = 0 and $g_n = 2$ for n $\geq$ 1. Integrating Eq.(4) and using the relation $n_e = N_{phase}$ gives an expression $E_F(e) \approx 77(B/B_{cr})^{1/4} MeV$ when B$\gg B_{cr}$. Once the energies of electrons are higher than the neutron Fermi kinetic energy (~60MeV), the process of electron capture (EC) will happen in a magnetar. The outgoing high-energy EC neutrons can easily destroy the $^3p_2$ Cooper pairs through the nuclear force interaction. This reaction is expressed as $n+(n\uparrow+n\uparrow)\to n+n+n$. When one Cooper pair is destroyed, the magnetic energy $2\mu_n B$ would be released and transformed into thermal energy, then can be radiated as soft X-rays, $kT \sim \mu_n B = 10 B_{15} keV$. The total magnetic energy of $^3p_2$ Cooper pairs can be estimated as $E = \frac{1}{2} q N_A m(^3p_2) \times 2\mu_n B \approx 1\times 10^{47} B_{15} \frac{m(^3p_2)}{0.1 m_{sun}}$ ergs. This energy can maintain over $10^{4-6}$ yrs for $L_X \sim 10^{34}-10^{36} ergs/s$ of a magnetar. According to the above ideas, we calculate the theoretical luminosities of magnetars, and compare our results with observations of magnetars. Apart from 1E 2259 and 4U 0142 in the upper left corner of Fig.1, for most magnetars, their soft X-ray luminosity observations are basically consistent with the theoretical curve. However, recent observations show that there could exist accretion disks around some magnetars[9], so obvious deviations from the curve seen in Fig.1 may produce.

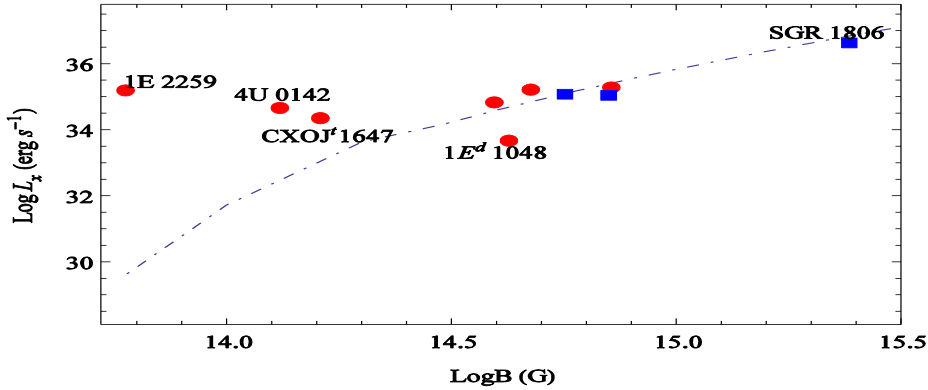

**Figure.1**

**Acknowledgements:** This work is supported by National Basic Research Program of China (973 Program 2009CB824800), the Key Directional Project of CAS and NSFC ( No 10173020,10673021, 10773005, 10778631 and 10903019).